\documentclass[twocolumn,10pt]{IEEEtran}
\usepackage[T1]{fontenc}
\usepackage[latin9]{inputenc}
\usepackage{amsmath}
\usepackage{amssymb}
\usepackage{amsthm}
\usepackage{graphicx}
\usepackage{comment}
\usepackage{booktabs}
\usepackage{xcolor}
\usepackage[ruled,vlined,linesnumbered]{algorithm2e}
\usepackage[acronym]{glossaries}
\glsdisablehyper
\newacronym{mimo}{MIMO}{multiple-input multiple-output}
\newacronym{miso}{MISO}{multiple-input single-output}
\newacronym{siso}{SISO}{single-input single-output}
\newacronym{awgn}{AWGN}{additive white Gaussian noise}
\newacronym{ula}{ULA}{uniform linear array}
\newacronym{upa}{UPA}{uniform planar array}
\newacronym{los}{LoS}{line-of-sight}
\newacronym{nlos}{NLoS}{non-line-of-sight}
\newacronym{star-ris}{STAR-RIS}{simultaneously transmitting and reflecting RIS}
\newacronym{ios}{IOS}{intelligent omni-surface}
\newacronym{mrt}{MRT}{maximal ratio transmission}
\newacronym{rsma}{RSMA}{rate splitting multiple access}
\newacronym{dfrc}{DFRC}{dual-function radar-communication}
\newacronym{mec}{MEC}{mobile edge computing}

\begin{document}

\title{\huge Static Grouping Strategy Design for Beyond Diagonal Reconfigurable Intelligent Surfaces}

\author{Matteo Nerini,~\IEEEmembership{Student~Member,~IEEE},
        Shanpu Shen,~\IEEEmembership{Senior~Member,~IEEE},
        Bruno Clerckx,~\IEEEmembership{Fellow,~IEEE}

\thanks{M. Nerini and B. Clerckx are with the Department of Electrical and Electronic Engineering, Imperial College London, SW7 2AZ London, U.K. (e-mail: m.nerini20@imperial.ac.uk, b.clerckx@imperial.ac.uk).}
\thanks{S. Shen is with the Department of Electrical Engineering and Electronics, University of Liverpool, L69 3GJ Liverpool, U.K. (email: Shanpu.Shen@liverpool.ac.uk).}}

\maketitle

\begin{abstract}
Beyond diagonal reconfigurable intelligent surface (BD-RIS) extends conventional RIS through novel architectures, such as group-connected RIS, with scattering matrix not restricted to being diagonal.
However, it remains unexplored how to optimally group the elements in group-connected RISs to maximize the performance while maintaining a low-complexity circuit.
In this study, we propose and model BD-RIS with a static grouping strategy optimized based on the channel statistics.
After formulating the corresponding problems, we design the grouping in single- and multi-user systems.
Numerical results reveal the benefits of grouping optimization, i.e., up to 60\% sum rate improvement, especially in highly correlated channels.
\end{abstract}

\glsresetall


\section{Introduction}

Reconfigurable intelligent surface (RIS) is a promising, cost-effective technology that can enhance performance and coverage in wireless networks \cite{wu21}.
RISs are surfaces constituted by multiple reflecting elements, able to modify the amplitude and phase of incident electromagnetic waves.
By optimizing the reflection coefficients of these elements, RISs can drive the reflected wave toward the intended direction.
Conventionally, RISs have been developed by independently controlling each element through a reconfigurable impedance, resulting in RISs characterized by a diagonal scattering matrix, more commonly referred to as phase shift matrix \cite{she20}.

More recently, beyond diagonal RIS (BD-RIS), i.e., the set of generalized RIS architectures with scattering matrices not restricted to being diagonal \cite{li23-1}, has emerged to break through the limitation of diagonal scattering matrices by allowing the signal impinging on one RIS element to be reflected from other elements \cite{li22}.
BD-RIS has shown notable gains in improving performance in wireless systems, such as \gls{rsma} \cite{fan22}, \gls{dfrc} \cite{wan23}, and \gls{mec} systems \cite{mah23}.
In the framework of BD-RIS, the conventional RIS is categorized as single-connected RIS, and further extended to fully-connected RIS by connecting all the RIS elements through tunable impedance components \cite{she20}.
The fully-connected RIS is the most flexible BD-RIS architecture, achieving maximum performance at the cost of a high circuit complexity.

To achieve a good trade-off between performance and complexity, the group-connected RIS was proposed by arranging the RIS elements into groups and interconnecting only elements within the same group \cite{she20,ner23-2}.
Interestingly, single- and fully-connected RISs are special cases of group-connected RIS.
Group-connected RIS has been optimized in single-user \cite{san23,ner22} and multi-user systems \cite{fan23} considering continuous reflection coefficients, as well as discretized reflection coefficients \cite{ner21}, showing promising gains over single-connected RIS.
In \cite{li22-1,li22-2}, the group-connected architecture has been exploited to both reflect and transmit the incident electromagnetic wave, enabling full-space coverage, different from conventional RIS \cite{zha22}.

In previous works on group-connected RIS \cite{she20}-\cite{li22-2}, the RIS elements are grouped simply by collecting adjacent RIS elements into a group, without considering the grouping strategy optimization.
In \cite{li22-3}, a dynamic group strategy for BD-RIS supporting hybrid transmitting and reflecting mode has been proposed, where the groups are dynamically optimized in real-time on a per-channel realization basis.
However, to implement the dynamic group strategy optimization, the circuit requires additional switches, which leads to higher circuit complexity compared to conventional group-connected RIS \cite{li22-3}.
Therefore, it is worthwhile to consider the design of a static grouping strategy, which does not require additional switches and control overhead, to enhance the rate achievable with group-connected RIS while maintaining the same circuit complexity.
To that end, in this study, we investigate how to effectively group the RIS elements of group-connected RIS to maximize the achievable rate while not increasing the circuit complexity.
Our contributions are outlined as follows.

\emph{First}, we provide the model of group-connected RIS with an optimized static grouping strategy, designed based on the channel statistics.
\emph{Second}, we formalize the corresponding optimization problems for single- and multi-user \gls{miso} systems and propose novel and efficient algorithms for the grouping strategy optimization.
\emph{Third}, we present numerical results to assess the gains derived from optimizing the grouping strategy, which can be of up to 60\% sum rate improvement in highly correlated channels.

\section{BD-RIS Model}

Consider an $N$-element BD-RIS, modeled as $N$ antenna elements connected to an $N$-port reconfigurable impedance network, with scattering matrix $\boldsymbol{\Theta}\in\mathbb{C}^{N\times N}$.
To maximize the power reflected by the BD-RIS, we consider the $N$-port reconfigurable impedance network to be lossless.
Thus, it can be described through its purely imaginary impedance matrix $\mathbf{Z}=j\mathbf{X}\in\mathbb{C}^{N\times N}$, where $\mathbf{X}\in\mathbb{R}^{N\times N}$ is the reactance matrix.
According to network theory \cite{poz11}, $\mathbf{X}$ and $\boldsymbol{\Theta}$ are related by
\begin{equation}
\boldsymbol{\Theta}=\left(j\mathbf{X}+Z_{0}\mathbf{I}\right)^{-1}\left(j\mathbf{X}-Z_{0}\mathbf{I}\right),\label{eq:T(X)}
\end{equation}
where $Z_0$ denotes the reference impedance according to which the scattering parameters are computed, typically set to $50$ $\Omega$.

In a group-connected RIS, the $N$ RIS elements are divided into $G$ groups, each including $N_G=N/G$ elements \cite{she20}.
Each RIS element is connected to ground and to all other elements in its group through a tunable impedance component, while there is no connection between groups\footnote{Our modeling is also valid for group-connected RIS with transmissive and reflective capabilities, as it has the same circuit topology as purely reflective group-connected RIS \cite{li22-1}.}.
In previous literature, the RIS elements have been grouped in sequential order according to their indices.
In this way, the $g$th group is composed of the $N_G$ elements indexed by
\begin{equation}
\mathcal{G}_g=\left\{\left(g-1\right)N_G+1,\ldots,gN_G\right\},\label{eq:NG}
\end{equation}
for $g=1,\ldots,G$.
As a result of this grouping, $\mathbf{X}$ and $\boldsymbol{\Theta}$ are block diagonal matrices in previous works \cite{she20}.

Differently from previous literature, in this study, we optimize the grouping strategy to additionally enhance the performance of group-connected RIS in terms of received power and sum rate in single- and multi-user systems, respectively.
The grouping strategy can be described by a permutation $\pi$ of the $N$ RIS elements such that the $g$th group is composed of the elements indexed by
\begin{equation}
\mathcal{G}_g=\left\{\pi\left(\left(g-1\right)N_G+1\right),\ldots,\pi\left(gN_G\right)\right\},
\end{equation}
for $g=1,\ldots,G$.
Accordingly, the reactance matrix is a permuted block diagonal matrix given by
\begin{equation}
\mathbf{X}=\mathbf{P}_{\pi}\bar{\mathbf{X}}\mathbf{P}_{\pi}^T,\label{eq:X}
\end{equation}
where $\mathbf{P}_{\pi}=[\mathbf{e}_{\pi(1)},\ldots,\mathbf{e}_{\pi(N)}]$ is the permutation matrix associated to $\pi$, with $\mathbf{e}_n\in\mathbb{R}^{N\times 1}$ denoting the vector with the $n$th entry being 1 and the others being 0, and $\bar{\mathbf{X}}\in\mathbb{R}^{N\times N}$ is a block diagonal matrix fulfilling
\begin{equation}
\bar{\mathbf{X}}=\mathrm{diag}\left(\bar{\mathbf{X}}_{1},\ldots,\bar{\mathbf{X}}_{G}\right),\:\bar{\mathbf{X}}_{g}=\bar{\mathbf{X}}_{g}^T,\:\forall g,
\end{equation}
with $\bar{\mathbf{X}}_{g}\in\mathbb{R}^{N_G\times N_G}$ being the reactance matrix of the $N_G$-port fully-connected reconfigurable impedance network for the $g$th group, which is symmetric for a reciprocal network.
As a consequence of \eqref{eq:T(X)} and \eqref{eq:X}, the scattering matrix writes as
\begin{equation}
\boldsymbol{\Theta}=\mathbf{P}_{\pi}\bar{\boldsymbol{\Theta}}\mathbf{P}_{\pi}^T,\label{eq:T}
\end{equation}
where $\bar{\boldsymbol{\Theta}}\in\mathbb{C}^{N\times N}$ is a block diagonal matrix satisfying
\begin{equation}
\bar{\boldsymbol{\Theta}}=\mathrm{diag}\left(\bar{\boldsymbol{\Theta}}_{1},\ldots,\bar{\boldsymbol{\Theta}}_{G}\right),\:\bar{\boldsymbol{\Theta}}_{g}=\bar{\boldsymbol{\Theta}}_{g}^T,\:\bar{\boldsymbol{\Theta}}_{g}^{H}\bar{\boldsymbol{\Theta}}_{g}=\mathbf{I},\:\forall g,
\end{equation}
with $\bar{\boldsymbol{\Theta}}_{g}\in\mathbb{C}^{N_G\times N_G}$ being the $g$th group scattering matrix, which is unitary for a lossless network.

To reconfigure $\boldsymbol{\Theta}$ given by \eqref{eq:T}, we optimize the grouping strategy $\pi$, impacting on $\mathbf{P}_\pi$, offline based on the channel statistics, i.e., mean and covariance matrix, while we optimize $\bar{\boldsymbol{\Theta}}$ online on a per-channel realization basis.
Since the grouping strategy is not reconfigured online, it is denoted as ``static''.
In the following, we formulate the corresponding optimization problems for single- and multi-user systems.


\section{Grouping Strategy Optimization for Single-User Systems}

Consider a single-user RIS-aided \gls{miso} system where the transmitter is equipped with $M$ antennas.
The channel $\mathbf{h}\in\mathbb{C}^{1\times M}$ between the transmitter and the receiver is expressed as $\mathbf{h}=\mathbf{h}_{R}\boldsymbol{\Theta}\mathbf{H}_{T}$, where $\mathbf{h}_{R}\in\mathbb{C}^{1\times N}$ and $\mathbf{H}_{T}\in\mathbb{C}^{N\times M}$ denote the channels from the RIS to the receiver and from the transmitter to the RIS, respectively.
The transmitted signal is $\mathbf{x}=\mathbf{w}s$, where $\mathbf{w}\in\mathbb{C}^{M\times 1}$ and $s\in\mathbb{C}$ are the precoding vector and data symbol.
The precoding vector is subject to $\Vert\mathbf{w}\Vert_2^2=1$ while the data symbol is subject to the transmit power constraint $\mathrm{E}[\vert s\vert^{2}]=P_T$.
Thus, the received signal is given by $y=\mathbf{h}_{R}\boldsymbol{\Theta}\mathbf{H}_{T}\mathbf{w}s+z$, where $z\sim\mathcal{CN}\left(0,\sigma_z^2\right)$ is the \gls{awgn}, with received signal power given by
\begin{equation}
P_R=P_T\left\vert\mathbf{h}_{R}\boldsymbol{\Theta}\mathbf{H}_{T}\mathbf{w}\right\vert^2,\label{eq:PR}
\end{equation}
which we want to maximize by optimizing $\boldsymbol{\Theta}$ and $\mathbf{w}$.
Observing that \eqref{eq:PR} is maximized when $\mathbf{w}$ is given by \gls{mrt}, i.e., $\mathbf{w}=\mathbf{h}^H/\Vert\mathbf{h}\Vert_2$, our problem reduces to maximize the channel gain $\Vert\mathbf{h}\Vert_2^2$ by optimizing $\boldsymbol{\Theta}$.

The resulting optimization problem can be formalized as a bi-level problem composed of a lower- and upper-level problem.
The lower-level problem, solved on a per-channel realization basis, is given by
\begin{align}
\underset{\bar{\boldsymbol{\Theta}}}{\mathsf{\mathrm{max}}}\;\;
&\left\Vert\mathbf{h}_{R}\mathbf{P}_\pi\bar{\boldsymbol{\Theta}}\mathbf{P}_\pi^T\mathbf{H}_{T}\right\Vert_2^2\label{eq:prob1-obj-su}\\
\mathsf{\mathrm{s.t.}}\;\;\;
&\bar{\boldsymbol{\Theta}}=\mathrm{diag}\left(\bar{\boldsymbol{\Theta}}_{1},\ldots,\bar{\boldsymbol{\Theta}}_{G}\right),\label{eq:prob1-c1-su}\\
&\bar{\boldsymbol{\Theta}}_{g}=\bar{\boldsymbol{\Theta}}_{g}^T,\:\bar{\boldsymbol{\Theta}}_{g}^{H}\bar{\boldsymbol{\Theta}}_{g}=\mathbf{I},\:\forall g,\label{eq:prob1-c2-su}
\end{align}
according to which the matrix $\bar{\boldsymbol{\Theta}}$ is optimized given a fixed permutation matrix $\mathbf{P}_\pi$ depending on the static grouping strategy $\pi$.
Besides, the upper-level problem, solved offline, optimizes $\pi$ based on a training set of $C$ channel realizations representative of the channel statistics.
This problem is formalized as
\begin{align}
\underset{\pi}{\mathsf{\mathrm{max}}}\;\;
&\frac{1}{C}\sum_{c=1}^C\left\Vert\mathbf{h}_{R}^{(c)}\mathbf{P}_\pi\bar{\boldsymbol{\Theta}}^{(c)}\mathbf{P}_\pi^T\mathbf{H}_{T}^{(c)}\right\Vert_2^2\label{eq:prob2-obj-su}\\
\mathsf{\mathrm{s.t.}}\;\;\;
&\mathbf{P}_\pi=[\mathbf{e}_{\pi(1)},\ldots,\mathbf{e}_{\pi(N)}],\label{eq:prob2-c1-su}\\
&\bar{\boldsymbol{\Theta}}^{(c)}\text{ solves \eqref{eq:prob1-obj-su}-\eqref{eq:prob1-c2-su}, }\forall c,\label{eq:prob2-c2-su}
\end{align}
where $\mathbf{h}_{R}^{(c)}$ and $\mathbf{H}_{T}^{(c)}$ are the $c$th channel realizations in the training set, for $c=1,\ldots,C$.
Remarkably, the bi-level problem \eqref{eq:prob1-obj-su}-\eqref{eq:prob2-c2-su} is hard to solve since it involves two nested optimization problems.

\subsection{Optimizing the Static Grouping Strategy $\pi$ Offline}

To efficiently optimize $\pi$ by solving \eqref{eq:prob2-obj-su}-\eqref{eq:prob2-c2-su}, we first remove the dependence on the variables $\{\bar{\boldsymbol{\Theta}}^{(c)}\}_{c=1}^{C}$ from the objective \eqref{eq:prob2-obj-su}.
To this end, we introduce $\bar{\mathbf{h}}_{R}^{(c)}=\mathbf{h}_{R}^{(c)}\mathbf{P}_\pi$ and $\bar{\mathbf{H}}_{T}^{(c)}=\mathbf{P}_\pi^T\mathbf{H}_{T}^{(c)}$, so that the objective $\Vert\mathbf{h}_{R}^{(c)}\mathbf{P}_\pi\bar{\boldsymbol{\Theta}}^{(c)}\mathbf{P}_\pi^T\mathbf{H}_{T}^{(c)}\Vert_2^2$ can be lower bounded by
\begin{align}
\left\Vert\bar{\mathbf{h}}_{R}^{(c)}\bar{\boldsymbol{\Theta}}^{(c)}\bar{\mathbf{H}}_{T}^{(c)}\right\Vert_2^2
&\geq\left\vert\bar{\mathbf{h}}_{R}^{(c)}\bar{\boldsymbol{\Theta}}^{(c)}\bar{\mathbf{H}}_{T}^{(c)}\mathbf{v}_T^{(c)}\right\vert^2\\
&=\left\Vert\mathbf{H}_{T}^{(c)}\right\Vert_2^2\left\vert\bar{\mathbf{h}}_{R}^{(c)}\bar{\boldsymbol{\Theta}}^{(c)}\mathbf{u}_T^{(c)}\right\vert^2,
\end{align}
where $\mathbf{v}_T^{(c)}\in\mathbb{C}^{M\times 1}$ and $\mathbf{u}_T^{(c)}\in\mathbb{C}^{N\times 1}$ are the dominant right and left singular vectors of $\bar{\mathbf{H}}_{T}^{(c)}$, respectively.
According to \cite{ner22}, we can find in closed-form the optimal $\bar{\boldsymbol{\Theta}}^{(c)}$ maximizing the terms $\vert\bar{\mathbf{h}}_{R}^{(c)}\bar{\boldsymbol{\Theta}}^{(c)}\mathbf{u}_T^{(c)}\vert^2$, denoted as $\bar{\boldsymbol{\Theta}}^{\star(c)}$, yielding 
\begin{equation}
\left\vert\bar{\mathbf{h}}_{R}^{(c)}\bar{\boldsymbol{\Theta}}^{\star(c)}\mathbf{u}_T^{(c)}\right\vert=\sum_{g=1}^G\left\Vert\left[\bar{\mathbf{h}}_{R}^{(c)}\right]_{\mathcal{G}_g}\right\Vert_2\left\Vert\left[\mathbf{u}_T^{(c)}\right]_{\mathcal{G}_g}\right\Vert_2,
\end{equation}
where $\mathcal{G}_g=\left\{\left(g-1\right)N_G+1,\ldots,gN_G\right\}$, for $g=1,\ldots,G$.
Thus, problem \eqref{eq:prob2-obj-su}-\eqref{eq:prob2-c2-su} can be simplified into
\begin{align}
\underset{\pi}{\mathsf{\mathrm{max}}}\;\;
&\frac{1}{C}\sum_{c=1}^C\left(\sum_{g=1}^G\left\Vert\left[\bar{\mathbf{h}}_{R}^{(c)}\right]_{\mathcal{G}_g}\right\Vert_2\left\Vert\left[\mathbf{u}_{T}^{(c)}\right]_{\mathcal{G}_g}\right\Vert_2\right)^2\label{eq:prob22-obj-su}\\
\mathsf{\mathrm{s.t.}}\;\;\;
&\mathbf{P}_\pi=[\mathbf{e}_{\pi(1)},\ldots,\mathbf{e}_{\pi(N)}],\label{eq:prob22-c1-su}
\end{align}
which no longer contains a nested optimization problem as the objective solely depends on the permutation matrix $\mathbf{P}_\pi$.

\begin{algorithm}[t]
\SetAlgoLined
\KwIn{$N_{G}, \{\mathcal{H}^{(c)}\}_{c=1}^{C}$}
\KwOut{The optimized grouping strategy $\pi^\star$}
$i\leftarrow 0$, $\pi_0(n)=n$, $\forall n$\;
\Repeat{$\pi_i=\pi_{i-1}$}{
$i\leftarrow i+1$\;
Update $\Pi_i$ as the set of all the grouping strategies obtainable by swapping two elements in $\pi_{i-1}$\;
Update $\pi_i$ by solving problem \eqref{eq:prob-obj-alg}-\eqref{eq:prob-c2-alg}\;
}
\KwRet{$\pi^\star\leftarrow \pi_i$};
\caption{Optimizing the grouping strategy.}
\label{alg:grouping-strategy}
\end{algorithm}

To solve problem \eqref{eq:prob22-obj-su}-\eqref{eq:prob22-c1-su} through exhaustive search has prohibitive complexity because of the high number of possible grouping strategies $\pi$.
Specifically, in RISs with $N$ elements grouped into $G$ groups, each containing $N_{G}$ elements, the number of possible grouping strategies $N_S$ is given by
\begin{equation}
N_{S}=\frac{1}{G!}\binom{N}{N_G}\binom{N-N_G}{N_G}\cdots\binom{N_G}{N_G}
=\frac{1}{G!}\frac{N!}{{\left(N_G!\right)}^G},
\end{equation}
growing with $N!$.
To decrease the search space, we solve problem \eqref{eq:prob22-obj-su}-\eqref{eq:prob22-c1-su} through a local search process, as shown in Alg.~\ref{alg:grouping-strategy}.
Specifically, we find the optimal grouping strategy $\pi^\star$ maximizing the objective \eqref{eq:prob22-obj-su} as follows.
The grouping strategy is initialized to the trivial permutation.
Thus, denoting as $\pi_0$ the initial grouping strategy, we have $\pi_0(n)=n$, for $n=1,\ldots,N$.
At the $i$th iteration, we generate all possible grouping strategies obtainable by swapping two elements in $\pi_{i-1}$, where the swapping operation consists of selecting two elements belonging to two different groups and assigning each of them to the group of the other.
The resulting set of grouping strategies is denoted as $\Pi_i$ and it is possible to show that its cardinality is $N(N-N_G)/2$.
Then, the objective \eqref{eq:prob22-obj-su} is computed for each grouping strategy in $\Pi_i$.
Finally, $\pi_i$ is given by the grouping strategy in $\{\Pi_i\cup \pi_{i-1}\}$ maximizing \eqref{eq:prob22-obj-su}, namely
\begin{align}
\underset{\pi}{\mathsf{\mathrm{max}}}\;\;
&\frac{1}{C}\sum_{c=1}^C\left(\sum_{g=1}^G\left\Vert\left[\bar{\mathbf{h}}_{R}^{(c)}\right]_{\mathcal{G}_g}\right\Vert_2\left\Vert\left[\mathbf{u}_{T}^{(c)}\right]_{\mathcal{G}_g}\right\Vert_2\right)^2\label{eq:prob-obj-alg}\\
\mathsf{\mathrm{s.t.}}\;\;\;
&\mathbf{P}_\pi=[\mathbf{e}_{\pi(1)},\ldots,\mathbf{e}_{\pi(N)}],\label{eq:prob-c1-alg}\\
&\pi\in\{\Pi_i\cup \pi_{i-1}\},\label{eq:prob-c2-alg}
\end{align}
which can be solved by an exhaustive search given the limited search space.
We update $\pi$ iteratively until the convergence is reached, i.e., when $\pi_i=\pi_{i-1}$.
Note that Alg.~\ref{alg:grouping-strategy} is ensured to converge by the following two properties.
First, the same grouping strategy is never selected in multiple iterations since \eqref{eq:prob22-obj-su} is strictly increasing over iterations.
Second, the total number of grouping strategies is limited.
The complexity of each iteration of Alg.~1 is driven by the complexity of solving problem \eqref{eq:prob-obj-alg}-\eqref{eq:prob-c2-alg}.
Since it is solved by an exhaustive search over the elements of $\Pi_i$, the complexity is equal to the cardinality of $\Pi_i$, i.e., $\mathcal{O}(N(N-N_G)/2)$.

\subsection{Optimizing the Scattering Matrix $\bar{\boldsymbol{\Theta}}$ Online}

When $\pi$ is fixed, we optimize $\bar{\boldsymbol{\Theta}}$ by solving problem \eqref{eq:prob1-obj-su}-\eqref{eq:prob1-c2-su} on a per-channel realization basis.
Introducing the equivalent channels $\bar{\mathbf{h}}_{R}=\mathbf{h}_{R}\mathbf{P}_\pi$ and $\bar{\mathbf{H}}_{T}=\mathbf{P}_\pi^T\mathbf{H}_{T}$, we solve \eqref{eq:prob1-obj-su}-\eqref{eq:prob1-c2-su} by alternatively optimizing $\bar{\boldsymbol{\Theta}}$ and the auxiliary variable $\mathbf{w}$ subject to $\Vert\mathbf{w}\Vert_2=1$ to maximize $\vert\bar{\mathbf{h}}_{R}\bar{\boldsymbol{\Theta}}\bar{\mathbf{H}}_{T}\mathbf{w}\vert^2$.
First, with fixed $\bar{\boldsymbol{\Theta}}$, we update $\mathbf{w}$ as $\mathbf{w}=(\bar{\mathbf{h}}_{R}\bar{\boldsymbol{\Theta}}\bar{\mathbf{H}}_{T})^H/\Vert\bar{\mathbf{h}}_{R}\bar{\boldsymbol{\Theta}}\bar{\mathbf{H}}_{T}\Vert_2$.
Second, with fixed $\mathbf{w}$, we update $\bar{\boldsymbol{\Theta}}$ through the global optimal solution provided in \cite{ner22}.
These two steps are iterated until the objective \eqref{eq:prob1-obj-su} converges.
The complexity of this online stage is driven by the complexity of the global optimal solution for $\bar{\boldsymbol{\Theta}}$ in \cite{ner22}, i.e., $\mathcal{O}(N_G^2N)$.

\section{Grouping Strategy Optimization for Multi-User Systems}

Consider a multi-user RIS-aided \gls{miso} system where the transmitter is equipped with $M$ antennas and there are $K$ single-antenna receivers.
The channel $\mathbf{h}_k\in\mathbb{C}^{1\times M}$ between the transmitter and the $k$th receiver can be expressed as $\mathbf{h}_k=\mathbf{h}_{R,k}\boldsymbol{\Theta}\mathbf{H}_{T}$, where $\mathbf{h}_{R,k}\in\mathbb{C}^{1\times N}$ and $\mathbf{H}_{T}\in\mathbb{C}^{N\times M}$ are the channels from the RIS to the $k$th receiver and from the transmitter to the RIS, respectively.
The transmitted signal is $\mathbf{x}=\sum_{k=1}^K\mathbf{w}_ks_k$, where $\mathbf{w}_k\in\mathbb{C}^{M\times 1}$ and $s_k\in\mathbb{C}$ are the precoding vector and data symbol for the $k$th receiver, which are subject to $\sum_{k=1}^K\left\Vert\mathbf{w}_k\right\Vert_2^2=1$ and $\mathrm{E}[\vert s_k\vert^{2}]=P_T$.
Thus, the signal at the $k$th receiver can be expressed as $y_k=\mathbf{h}_k\sum_{i}\mathbf{w}_is_i+z_k$, where $z_k\sim\mathcal{CN}\left(0,\sigma_z^2\right)$ is the \gls{awgn} at the $k$th receiver, yielding a sum rate given by
\begin{equation}
S_R=\sum_{k=1}^K\log\left(1+\frac{\left\vert\mathbf{h}_k\mathbf{w}_k\right\vert^2}{\sum_{i\neq k}\left\vert\mathbf{h}_k\mathbf{w}_i\right\vert^2+\sigma_z^2}\right).\label{eq:SR}
\end{equation}
To maximize \eqref{eq:SR} by jointly optimizing $\boldsymbol{\Theta}$ and $\mathbf{w}_1,\ldots,\mathbf{w}_K$ is a hard problem given the non-convex objective function.
Thus, we optimize the BD-RIS and the precoding vectors in two different stages, as also adopted in \cite{fan23}.
First, the BD-RIS is optimized to maximize the sum of the channel gains of all users $\sum_k\Vert\mathbf{h}_{k}\Vert_2^2=\Vert\mathbf{H}\Vert_F^2$, where $\mathbf{H}=[\mathbf{h}_{1}^T,\ldots,\mathbf{h}_{K}^T]^T\in\mathbb{C}^{K\times M}$.
Second, the precoding vectors $\mathbf{w}_1,\ldots,\mathbf{w}_K$ are designed through zero-forcing beamforming.

The resulting optimization problem can be formalized as follows.
The lower-level optimization problem is given by
\begin{align}
\underset{\bar{\boldsymbol{\Theta}}}{\mathsf{\mathrm{max}}}\;\;
&\left\Vert\mathbf{H}_{R}\mathbf{P}_\pi\bar{\boldsymbol{\Theta}}\mathbf{P}_\pi^T\mathbf{H}_{T}\right\Vert_F^2\label{eq:prob1-obj-mu}\\
\mathsf{\mathrm{s.t.}}\;\;\;
&\bar{\boldsymbol{\Theta}}=\mathrm{diag}\left(\bar{\boldsymbol{\Theta}}_{1},\ldots,\bar{\boldsymbol{\Theta}}_{G}\right),\label{eq:prob1-c1-mu}\\
&\bar{\boldsymbol{\Theta}}_{g}=\bar{\boldsymbol{\Theta}}_{g}^T,\:\bar{\boldsymbol{\Theta}}_{g}^{H}\bar{\boldsymbol{\Theta}}_{g}=\mathbf{I},\:\forall g,\label{eq:prob1-c2-mu}
\end{align}
where we introduced $\mathbf{H}_R=[\mathbf{h}_{R,1}^T,\ldots,\mathbf{h}_{R,K}^T]^T\in\mathbb{C}^{K\times N}$ and $\mathbf{P}_\pi$ is fixed.
In addition, the upper-level problem writes as
\begin{align}
\underset{\pi}{\mathsf{\mathrm{max}}}\;\;
&\frac{1}{C}\sum_{c=1}^C\left\Vert\mathbf{H}_{R}^{(c)}\mathbf{P}_\pi\bar{\boldsymbol{\Theta}}^{(c)}\mathbf{P}_\pi^T\mathbf{H}_{T}^{(c)}\right\Vert_F^2\label{eq:prob2-obj-mu}\\
\mathsf{\mathrm{s.t.}}\;\;\;
&\mathbf{P}_\pi=[\mathbf{e}_{\pi(1)},\ldots,\mathbf{e}_{\pi(N)}],\label{eq:prob2-c1-mu}\\
&\bar{\boldsymbol{\Theta}}^{(c)}\text{ solves \eqref{eq:prob1-obj-mu}-\eqref{eq:prob1-c2-mu}, }\forall c,\label{eq:prob2-c2-mu}
\end{align}
which results in a challenging bi-level problem.

\subsection{Optimizing the Static Grouping Strategy $\pi$ Offline}

Similarly to the single-user case, we optimize $\pi$ through \eqref{eq:prob2-obj-mu}-\eqref{eq:prob2-c2-mu} by removing the dependence on the variables $\{\bar{\boldsymbol{\Theta}}^{(c)}\}_{c=1}^{C}$ from the objective \eqref{eq:prob2-obj-mu}.
To this end, introducing $\bar{\mathbf{H}}_{R}^{(c)}=\mathbf{H}_{R}^{(c)}\mathbf{P}_\pi$, we can lower bound the objective $\Vert\mathbf{H}_{R}^{(c)}\mathbf{P}_\pi\bar{\boldsymbol{\Theta}}^{(c)}\mathbf{P}_\pi^T\mathbf{H}_{T}^{(c)}\Vert_F^2$ as
\begin{align}
\left\Vert\bar{\mathbf{H}}_{R}^{(c)}\bar{\boldsymbol{\Theta}}^{(c)}\bar{\mathbf{H}}_{T}^{(c)}\right\Vert_F^2
&\geq\left\Vert\bar{\mathbf{H}}_{R}^{(c)}\bar{\boldsymbol{\Theta}}^{(c)}\bar{\mathbf{H}}_{T}^{(c)}\right\Vert_2^2\\
&\geq\left\Vert\mathbf{H}_{R}^{(c)}\right\Vert_2^2\left\Vert\mathbf{H}_{T}^{(c)}\right\Vert_2^2\left\vert\mathbf{v}_{R}^{(c)H}\bar{\boldsymbol{\Theta}}^{(c)}\mathbf{u}_T^{(c)}\right\vert^2,
\end{align}
where $\mathbf{u}_R^{(c)}\in\mathbb{C}^{K\times 1}$ and $\mathbf{v}_R^{(c)}\in\mathbb{C}^{N\times 1}$ are the dominant left and right singular vectors of $\bar{\mathbf{H}}_{R}^{(c)}$.
The terms $\vert\mathbf{v}_{R}^{(c)H}\bar{\boldsymbol{\Theta}}^{(c)}\mathbf{u}_T^{(c)}\vert^2$ can be maximized in closed-form \cite{ner22}, giving
\begin{equation}
\left\vert\mathbf{v}_{R}^{(c)H}\bar{\boldsymbol{\Theta}}^{\star(c)}\mathbf{u}_T^{(c)}\right\vert=\sum_{g=1}^G\left\Vert\left[\mathbf{v}_{R}^{(c)}\right]_{\mathcal{G}_g}\right\Vert_2\left\Vert\left[\mathbf{u}_T^{(c)}\right]_{\mathcal{G}_g}\right\Vert_2,
\end{equation}
where $\mathcal{G}_g=\left\{\left(g-1\right)N_G+1,\ldots,gN_G\right\}$, for $g=1,\ldots,G$.
Consequently, problem \eqref{eq:prob2-obj-mu}-\eqref{eq:prob2-c2-mu} can be rewritten as
\begin{align}
\underset{\pi}{\mathsf{\mathrm{max}}}\;\;
&\frac{1}{C}\sum_{c=1}^C\left(\sum_{g=1}^G\left\Vert\left[\mathbf{v}_{R}^{(c)}\right]_{\mathcal{G}_g}\right\Vert_2\left\Vert\left[\mathbf{u}_{T}^{(c)}\right]_{\mathcal{G}_g}\right\Vert_2\right)^2\label{eq:prob22-obj-mu}\\
\mathsf{\mathrm{s.t.}}\;\;\;
&\mathbf{P}_\pi=[\mathbf{e}_{\pi(1)},\ldots,\mathbf{e}_{\pi(N)}],\label{eq:prob22-c1-mu}
\end{align}
in which the dependence on $\{\bar{\boldsymbol{\Theta}}^{(c)}\}_{c=1}^{C}$ has been removed.
Since the objective in \eqref{eq:prob22-obj-mu}-\eqref{eq:prob22-c1-mu} solely depends on the permutation matrix $\mathbf{P}_\pi$, it can be solved through Alg.~\ref{alg:grouping-strategy} by replacing \eqref{eq:prob22-obj-mu} into the objective of problem \eqref{eq:prob-obj-alg}-\eqref{eq:prob-c2-alg}.

\subsection{Optimizing the Scattering Matrix $\bar{\boldsymbol{\Theta}}$ Online}

Once $\pi$ has been fixed offline, $\bar{\boldsymbol{\Theta}}$ is reconfigured by solving problem \eqref{eq:prob1-obj-mu}-\eqref{eq:prob1-c2-mu} for each given channel realization.
Constraints \eqref{eq:prob1-c1-mu}-\eqref{eq:prob1-c2-mu} indicate that $\bar{\boldsymbol{\Theta}}$ is a block diagonal matrix with each block being a complex symmetric unitary matrix, which complicates the optimization.
Thus, the relationship between $\bar{\boldsymbol{\Theta}}$ and $\bar{\mathbf{X}}$ deriving from \eqref{eq:T(X)} can be exploited to equivalently reformulate \eqref{eq:prob1-obj-mu}-\eqref{eq:prob1-c2-mu} as
\begin{align}
\underset{\bar{\mathbf{X}}_{g}}{\mathsf{\mathrm{max}}}\;\;
&\left\Vert\bar{\mathbf{H}}_{R}\bar{\boldsymbol{\Theta}}\bar{\mathbf{H}}_{T}\right\Vert_F^2\label{eq:prob11-obj}\\
\mathsf{\mathrm{s.t.}}\;\;\;
&\bar{\boldsymbol{\Theta}}=\mathrm{diag}\left(\bar{\boldsymbol{\Theta}}_{1},\ldots,\bar{\boldsymbol{\Theta}}_{G}\right),\label{eq:prob11-c1}\\
&\bar{\boldsymbol{\Theta}}_{g}=\left(j\bar{\mathbf{X}}_{g}+Z_{0}\mathbf{I}\right)^{-1}\left(j\bar{\mathbf{X}}_{g}-Z_{0}\mathbf{I}\right),\:\forall g,\label{eq:prob11-c2}\\
&\bar{\mathbf{X}}_{g}=\bar{\mathbf{X}}_{g}^{T},\:\forall g,\label{eq:prob11-c3}
\end{align}
where we introduced the equivalent channels $\bar{\mathbf{H}}_{R}=\mathbf{H}_{R}\mathbf{P}_\pi$ and $\bar{\mathbf{H}}_{T}=\mathbf{P}_\pi^T\mathbf{H}_{T}$, which can be directly transformed into an unconstrained problem.
More precisely, exploiting the constraints \eqref{eq:prob11-c1}-\eqref{eq:prob11-c3}, the objective \eqref{eq:prob11-obj} can be expressed as a function of $\bar{\mathbf{X}}_{1},\ldots,\bar{\mathbf{X}}_{G}$.
Since $\bar{\mathbf{X}}_{g}$ is an arbitrary $N_G\times N_G$ real symmetric matrix, $\bar{\mathbf{X}}_{g}$ is an unconstrained function of the $N_G(N_G+1)/2$ entries in its upper triangular part.
Thus, problem \eqref{eq:prob11-obj}-\eqref{eq:prob11-c3} is an unconstrained problem in the variables $[\bar{\mathbf{X}}_{g}]_{i,j}$, with $i\leq j$, $\forall g$, and can be solved by using the quasi-Newton method to find the optimal upper triangular part of each block $\bar{\mathbf{X}}_{g}$ without any constraints.
The complexity of this online stage is given by the complexity of the quasi-Newton algorithm, i.e., $\mathcal{O}(N^2(N_G+1)^2/4)$ \cite{ner22}.

\section{Performance Evaluation}

We now evaluate the performance of group-connected RIS with an optimized grouping strategy.
The transmitter, the RIS, and the receiver(s) are located at $(0,0)$, $(50,2)$, and $(52,0)$ in meters (m), respectively.
The transmitter is equipped with an \gls{ula} composed of $M=4$ antennas, while the RIS is an \gls{upa} composed of $N_H\times N_V$ antennas, with $N_V=8$ and $N_HN_V=N$.
The path loss of the channels is given by the distance-dependent model $L_{i}(d_{i})=L_{0}(d_{i})^{-\alpha_{i}}$, where $L_{0}$ is the reference path loss at distance $1$~m, $d_{i}$ is the distance, and $\alpha_{i}$ is the path loss exponent for $i\in\{R,T\}$.
We set $L_{0}=-30$~dB, $\alpha_{R}=2.8$, and $\alpha_{T}=2$.
The small-scale fading effects are modeled as correlated Rayleigh fading, i.e., $\mathbf{h}_{R,k}\sim\mathcal{CN}\left(\boldsymbol{0},L_R\mathbf{R}_{RIS}\right)$, for $k=1,\ldots,K$, and $\text{vec}(\mathbf{H}_{T})\sim\mathcal{CN}\left(\boldsymbol{0},L_T\mathbf{R}_{T}\right)$.
The covariance matrix $\mathbf{R}_{RIS}\in\mathbb{R}^{N\times N}$ is given by the Kronecker correlation model for \glspl{upa} $\mathbf{R}_{RIS}=\mathbf{R}_{H}\otimes\mathbf{R}_{V}$, where $\mathbf{R}_{H}\in\mathbb{R}^{N_H\times N_H}$ and $\mathbf{R}_{V}\in\mathbb{R}^{N_V\times N_V}$ are defined through the exponential correlation model as $[\mathbf{R}_{H}]_{i,j}=\rho^{\vert i-j\vert}$ and $[\mathbf{R}_{V}]_{i,j}=\rho^{\vert i-j\vert}$, with $\rho$ being the correlation coefficient.
Besides, $\mathbf{R}_{T}$ is given by the Kronecker correlation model for \gls{mimo} channels $\mathbf{R}_{T}=\mathbf{R}_{RIS}\otimes\mathbf{R}_{TX}$, where $\mathbf{R}_{TX}\in\mathbb{R}^{M\times M}$ is defined as $[\mathbf{R}_{TX}]_{i,j}=\rho^{\vert i-j\vert}$.
We consider mildly correlated channels by setting $\rho=0.6$ and highly correlated channels with $\rho=0.8$.

\begin{figure}[t]
\centering
\includegraphics[width=0.24\textwidth]{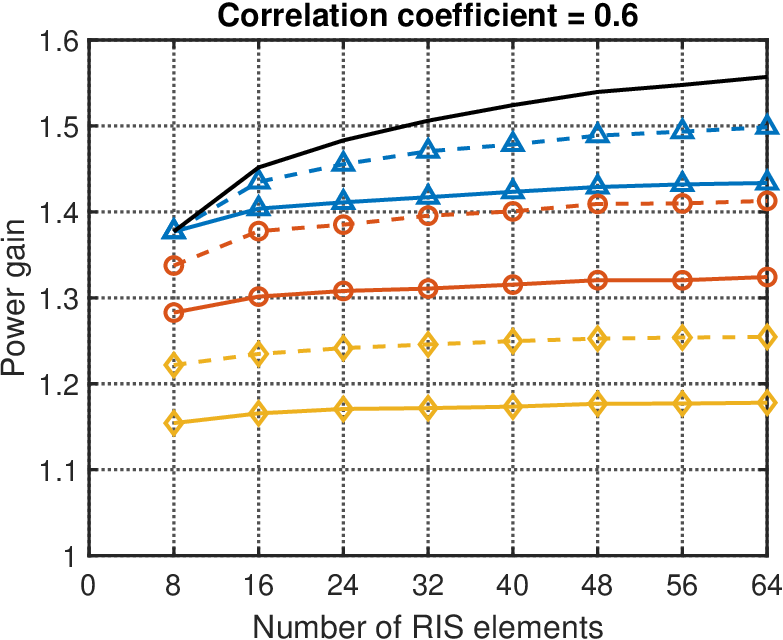}
\includegraphics[width=0.24\textwidth]{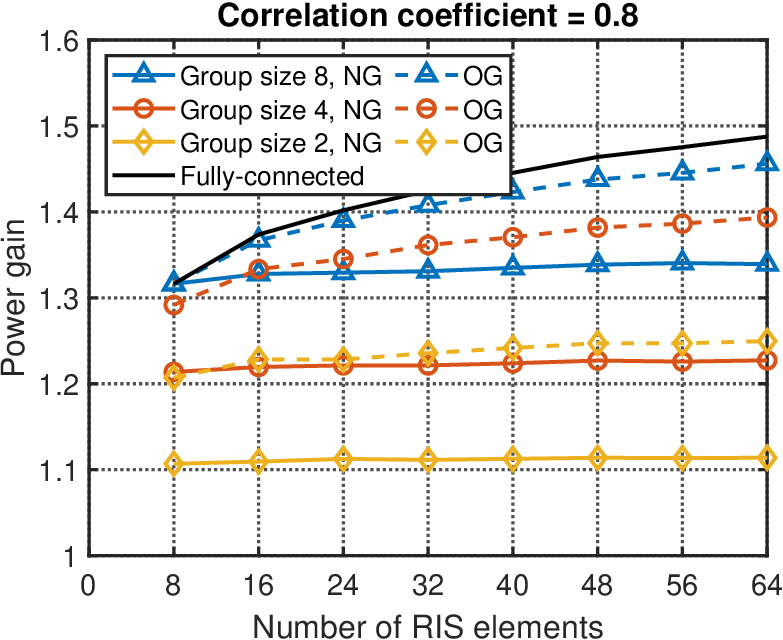}
\caption{Power gain in single-user systems aided by fully- and group-connected RISs with non-optimized grouping ``NG'' and optimized grouping ``OG''.}
\label{fig:power-gain}
\end{figure}
\begin{figure}[t]
\centering
\includegraphics[width=0.44\textwidth]{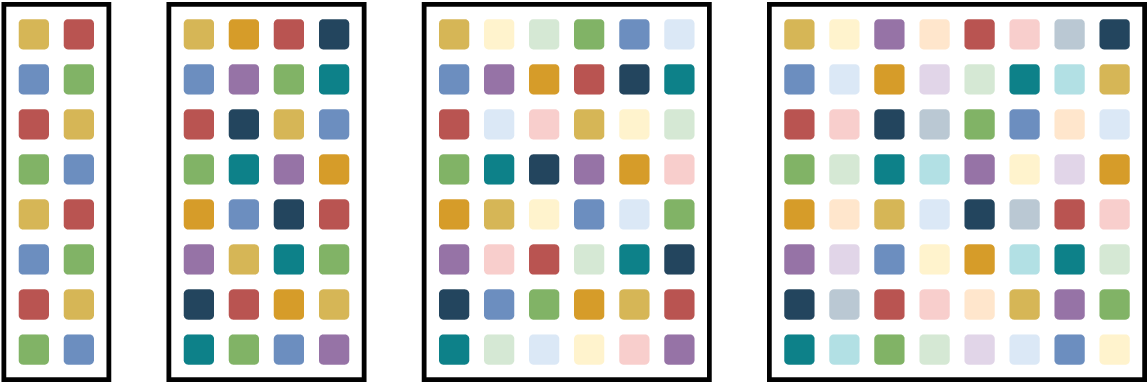}
\caption{Optimized grouping for RISs with $2\times8$, $4\times8$, $6\times8$, and $8\times8$ elements and group size 4.}
\label{fig:grouping}
\end{figure}

In Fig.~\ref{fig:power-gain}, we investigate the impact of the grouping strategy optimization in single-user systems.
To this end, we report the power gains of fully- and group-connected RISs over single-connected RIS, defined as $\mathcal{G}^{\mathrm{Fully}}=P_{R}^{\mathrm{Fully}}/P_{R}^{\mathrm{Single}}$ and $\mathcal{G}^{\mathrm{Group}}=P_{R}^{\mathrm{Group}}/P_{R}^{\mathrm{Single}}$, respectively, where $P_{R}^{\mathrm{Fully}}$, $P_{R}^{\mathrm{Single}}$, and $P_{R}^{\mathrm{Group}}$ are the received signal power given by \eqref{eq:PR} of fully-, single-, and group-connected RISs, respectively.
As a baseline, we consider the non-optimized grouping strategy given by \eqref{eq:NG}, obtained by grouping adjacent RIS elements, as in previous works \cite{she20}.
We observe that an optimized grouping strategy can significantly improve the performance of group-connected RIS, for any group size, particularly in the presence of highly correlated channels.
With $\rho=0.8$, the power gain is improved by up to 13\%, when $N=64$ and $N_G=4$.

In Fig.~\ref{fig:grouping}, we illustrate the resulting optimized grouping strategy in single-user systems with $\rho=0.8$, for RISs with group size $N_G=4$, where RIS elements in the same group have the same color.
Interestingly, the optimized grouping tends to maximize the distance between RIS elements grouped together.
Note that an optimized static grouping strategy does not require additional hardware and online optimization complexity, different from a dynamic grouping strategy \cite{li22-3}.

\begin{figure}[t]
\centering
\includegraphics[width=0.24\textwidth]{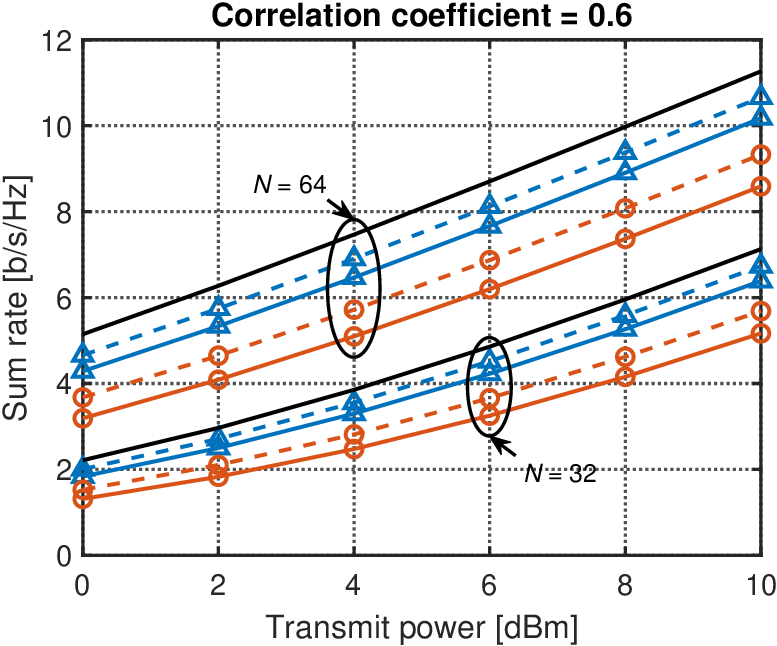}
\includegraphics[width=0.24\textwidth]{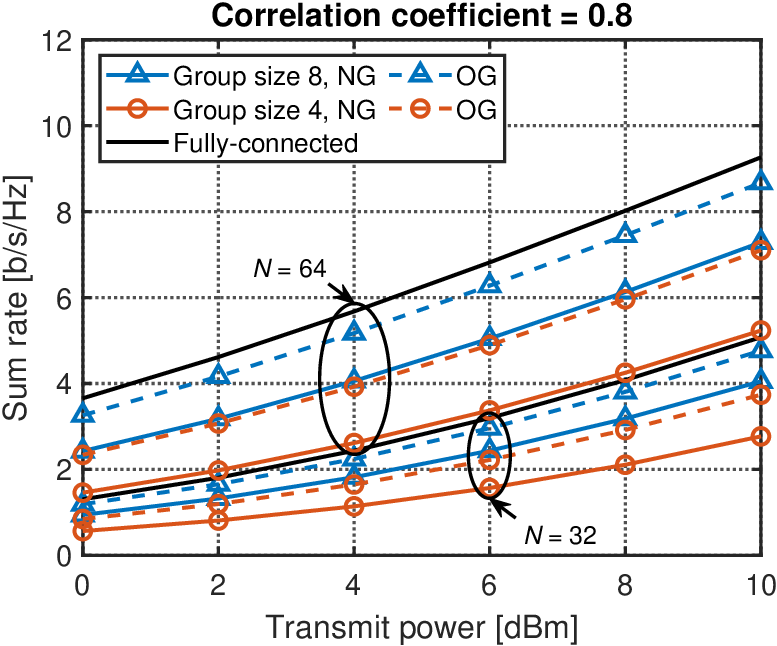}
\caption{Sum rate in multi-user systems aided by fully- and group-connected RISs with non-optimized grouping ``NG'' and optimized grouping ``OG''.}
\label{fig:sum-rate}
\end{figure}

In Fig.~\ref{fig:sum-rate}, we report the sum rate given by \eqref{eq:SR} obtained by group-connected RISs in multi-user systems with $K=2$ users and $\sigma_z^2=-80$~dBm.
As expected, the sum rate increases with the number of RIS elements $N$ and the group size $N_G$.
Besides, we notice that optimizing the grouping strategy can also visibly contribute to improving the sum rate by maintaining the same circuit complexity.
Also in multi-user systems, the grouping strategy especially impacts the performance in highly correlated channels.
Specifically, with $\rho=0.8$, the sum rate is improved by up to 60\% in multi-user systems with $N=64$ and $N_G=4$.

\section{Conclusion}

We address the problem of optimally grouping the RIS elements in group-connected RIS based on the channel statistics, to improve the performance of BD-RIS while not increasing the circuit complexity.
To this end, we show how to optimize the grouping based on the channel statistics (offline) and the tunable impedance components on a per-channel realization basis (online) in single- and multi-user systems.
Numerical results show the gain of optimizing the static grouping strategy, especially in the presence of highly correlated channels.

\bibliographystyle{IEEEtran}
\bibliography{IEEEabrv,main}
\end{document}